\documentclass[3p,times,twocolumn,authoryear,nopreprintline]{elsarticle}

\usepackage{sty}
\usepackage{listings}
\usepackage{shortcuts}
\usepackage{multirow}
\usepackage{amsmath,amssymb,amsfonts}
\usepackage{algorithmic}
\usepackage{graphicx}
\usepackage{textcomp}
\usepackage{xcolor}

\newcommand{\si}[1]{\small\texttt{#1}\normalsize}
\newcommand{\ti}[1]{\texttt{#1}}
\newcommand{\mr}[2]{\multirow{#1}{*}{#2}}

\lstdefinestyle{cstyle}{
language=c,
basicstyle=\ttfamily\bfseries\scriptsize,
  morekeywords={virtualinvoke},
  keywordstyle=\color{blue},
  ndkeywordstyle=\color{red},
  commentstyle=\color{dkred},
  keepspaces=true,              % very important to preserve space after ' in footnote
  numbers=left,
  breaklines=true,
  numberstyle=\ttfamily\footnotesize\color{gray},
  stepnumber=1,
  numbersep=10pt,
  backgroundcolor=\color{white},
  tabsize=4,
  showspaces=false,
  showstringspaces=false,
  xleftmargin=.23in,
  captionpos=b,
  escapeinside={(?}{?)},
}

\lstdefinestyle{comby}{
  basicstyle=\footnotesize\ttfamily,
  numbers=none,
  escapeinside={@?}{?@},
  classoffset=0,
  keywordstyle=\color{blue},
  classoffset=1,
  keywordstyle=\color{orange},
  morekeywords={val,let,fun,match,with,rec,raise,return,type},
  classoffset=2,
  keywordstyle=\color{purple},
  morekeywords={list,string,unit,error},
  classoffset=3,
  morecomment=[s]{(**}{*)},
  moredelim=[is][\color{blue}]{@X}{X@},
  moredelim=[is][\color{purple}]{@Y}{Y@},
  escapeinside={(?}{?)},
}

\usepackage{amssymb}
\usepackage[figuresright]{rotating}

\begin{document}

\begin{frontmatter}

%% Title, authors and addresses

%% use the tnoteref command within \title for footnotes;
%% use the tnotetext command for the associated footnote;
%% use the fnref command within \author or \address for footnotes;
%% use the fntext command for the associated footnote;
%% use the corref command within \author for corresponding author footnotes;
%% use the cortext command for the associated footnote;
%% use the ead command for the email address,
%% and the form \ead[url] for the home page:
%%
%% \title{Title\tnoteref{label1}}
%% \tnotetext[label1]{}
%% \author{Name\corref{cor1}\fnref{label2}}
%% \ead{email address}
%% \ead[url]{home page}
%% \fntext[label2]{}
%% \cortext[cor1]{}
%% \address{Address\fnref{label3}}
%% \fntext[label3]{}

%\dochead{}
%% Use \dochead if there is an article header, e.g. \dochead{Short communication}
%% \dochead can also be used to include a conference title, if directed by the editors
%% e.g. \dochead{17th International Conference on Dynamical Processes in Excited States of Solids}

\title{Towards Fully Declarative Program Analysis via Source Code Transformation}

%% use optional labels to link authors explicitly to addresses:
%% \author[label1,label2]{<author name>}
%% \address[label1]{<address>}
%% \address[label2]{<address>}

\author{Rijnard van Tonder}

\address{Sourcegraph Inc., 981 Mission St, San Francisco, CA 94103, USA}

\begin{abstract}
Advances in logic programming and increasing industrial uptake of
Datalog-inspired approaches demonstrate the emerging need to express
powerful code analyses more easily. Declarative program analysis frameworks
(e.g., using logic programming like Datalog) significantly ease defining
analyses compared to imperative implementations. However, the declarative
benefits of these frameworks only materialize after parsing and translating source
code to generate facts. Fact generation remains a non-declarative precursor to
analysis where imperative implementations first parse and interpret program
structures (e.g., abstract syntax trees and control-flow graphs). The procedure
of fact generation thus remains opaque and difficult for non-experts to
understand or modify. We present a new perspective on this analysis workflow by
proposing \emph{declarative fact generation} to ease specification and
exploration of lightweight declarative analyses. Our approach demonstrates the
first venture towards fully declarative analysis specification across multiple
languages. The key idea is to translate source code \emph{directly} to Datalog
facts in the analysis domain using declarative syntax transformation. We then
reuse existing Datalog analyses over generated facts, yielding an end-to-end
declarative pipeline. As a first approximation we pursue a syntax-driven
approach and demonstrate the feasibility of generating and using lightweight
versions of liveness and call graph reachability properties. We then discuss the
workability of extending declarative fact generation to also incorporate
semantic information.
\end{abstract}
\begin{keyword}
%% keywords here, in the form: keyword \sep keyword
%
%% PACS codes here, in the form: \PACS code \sep code
%
%% MSC codes here, in the form: \MSC code \sep code
%% or \MSC[2008] code \sep code (2000 is the default)
declarative programming \sep program analysis \sep program transformation \sep Datalog \sep software quality
\end{keyword}
\end{frontmatter}

\section{Introduction}
\label{sec:intro}

Advances in logic programming~\citep{logicblox} and increasing industrial uptake
of Datalog-inspired approaches (e.g., CodeQL~\citep{codeqlweb,codeql},
Glean~\citep{glean}) demonstrate the emerging need for more powerful program
analyses, specified simply and declaratively. Fact generation is the process of
translating source code to a program model of Datalog propositions (like
``variable $x$ is read on line $l$''). Such facts underpin declarative program
analyses that answer queries over program properties (e.g., liveness).
Fact generation is typically accomplished by imperative code that first parses the
input program according to a known language grammar.
Program terms are further resolved to the domain of relevant facts for the
analysis, which depends on the input language's syntax and semantics. While
it is straightforward to declaratively define a general liveness analysis (all
rules and propositions are succinctly specified a-priori), the act of
translating source code to facts incurs many degrees of unspoken complexity
(e.g., parsing language-specific constructs and resolving semantic properties
like types). This complexity invites up front \emph{imperative} (rather than
declarative) implementation that recurs per language. The implementation burden
contributes to deep, yet narrow support for a single language or small set of languages
(e.g., pointer analysis for Java~\citep{BravenboerS09}).
Current approaches lack convenient, declarative
processes for translating 
source code to logic facts in the analysis domain.
This impedes prototyping and developing fully declarative analysis pipelines for lightweight and language-general use cases.

\begin{figure}[t!]
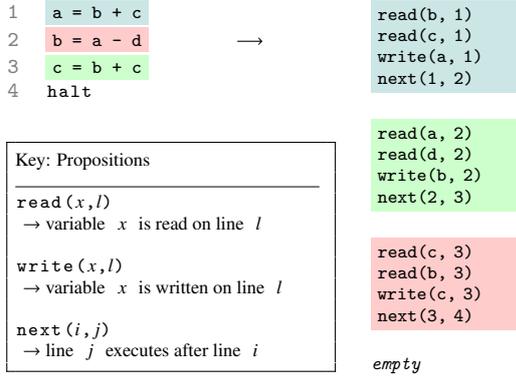

%\hspace*{2em}
\begin{subfigure}[t]{.6\columnwidth}
\begin{lstlisting}[basicstyle=\scriptsize\ttfamily,numbers=left,stepnumber=1,escapechar=!,xleftmargin=28pt]
!\tikz[baseline]{ \node[fill=blue!20,anchor=base] (one) {a = b + c}; }!
!\tikz[baseline]{ \node[fill=red!20,anchor=base] (two) {b = a - d}; }! !\hspace*{1cm}$\longrightarrow$!
!\tikz[baseline]{ \node[fill=green!20,anchor=base] (three) {c = b + c}; }!
halt
\end{lstlisting}
\vspace{.4em}
\begin{lstlisting}[basicstyle=\scriptsize\ttfamily,frame=single,numbers=none]
(?\textrm{Key: Propositions}?)
(?\rule{4cm}{0.1pt}?)
read((?$x$?),(?$l$?))
(?\hspace*{.1cm}$\rightarrow$\textrm{ variable}?) (?$x$?) (?\textrm{is read on line}?) (?$l$?)

write((?$x$?),(?$l$?))
(?\hspace*{.1cm}$\rightarrow$\textrm{ variable}?) (?$x$?) (?\textrm{is written on line}?) (?$l$?)

next((?$i$?),(?$j$?))
(?\hspace*{.1cm}$\rightarrow$\textrm{ line}?) (?$j$?) (?\textrm{executes after line}?) (?$i$?)
\end{lstlisting}
\end{subfigure}%
\begin{subfigure}[t]{.5\columnwidth}
\begin{lstlisting}[basicstyle=\scriptsize\ttfamily,numbers=none]
(?\tikz[baseline]{ \node[fill=blue!20,anchor=base,text width=1.8cm] (one) {read(b, 1)\\read(c, 1)\\write(a, 1)\\next(1, 2)}; }?)

(?\tikz[baseline]{ \node[fill=green!20,anchor=base,text width=1.8cm] (two) {read(a, 2)\\read(d, 2)\\write(b, 2)\\next(2, 3)}; }?)

(?\tikz[baseline]{ \node[fill=red!20,anchor=base,text width=1.8cm] (three) {read(c, 3)\\read(b, 3)\\write(c, 3)\\next(3, 4)}; ?)

(?\emph{empty}?)
\end{lstlisting}
\end{subfigure}
\caption{Example translation of a simple arithmetic language to facts.}
\label{fig:example1}
\end{figure}

Our idea addresses the gap by proposing techniques for \emph{declarative
fact generation} by directly translating program source code to Datalog facts.
To give an example, consider Listing~\ref{fig:example1}
with a simple arithmetic language (left) and
corresponding propositions \si{read}, \si{write}, and \si{next} generated by
program statements (right).
In Datalog terminology, the propositions on the right establish the extensional
database (EDB), which are facts generated by translating the program. The intensional database (IDB) for our liveness analysis comprises the rule \si{live(}$x$, $l$\si{)}
which expresses that variable $x$ is live at line $l$. \\

Rules for computing the liveness of a variable $x$ at line $l$ is given by:

\footnotesize
\begin{align*}
\texttt{live(}x, l\texttt{)} & \leftarrow \texttt{read(}x, l\texttt{)} \\
\texttt{live(}x, l\texttt{)} & \leftarrow \texttt{live(}x, i\texttt{)}, \texttt{ next(}i, l\texttt{)}, \texttt{ }\neg \texttt{write(}x, l\texttt{)}
\end{align*}
\normalsize

That is, a variable $x$ is considered live at $l$ if it is read on line $l$, or
if it is transitively live and not overwritten on line $l$. This captures the
intuitive description that a variable is live if it is read on line $l$, or we
anticipate that it will be read at some point after $l$ in the program's
execution, before being potentially overwritten. As an admittedly naive
introduction, we might consider regular expressions as a crude way to generate
the facts from expressions for our liveness analysis, in lieu of a well-specified
parser:

\begin{figure}[ht]
\begin{subfigure}[t]{.58\columnwidth}
\begin{lstlisting}[basicstyle=\footnotesize\ttfamily,numbers=left,numbers=none,xleftmargin=6pt]
(?{\color{blue}(\verb|\w+|)}?) = (?{\color{blue}(\verb|\w+|)}?) \+ (?{\color{blue}(\verb|\w+|)}?) (?\hspace*{.22cm}$\longrightarrow$?)
\end{lstlisting}
\end{subfigure}%
\begin{subfigure}[t]{.40\columnwidth}
\begin{lstlisting}[basicstyle=\footnotesize\ttfamily,numbers=none]
read((?{\color{blue}\$2}?), (?{\color{purple}\verb|?|}?))
read((?{\color{blue}\$3}?), (?{\color{purple}\verb|?|}?))
write((?{\color{blue}\$1}?), (?{\color{purple}\verb|?|}?))
next((?{\color{blue}\verb|?|}?), (?{\color{purple}\verb|?|}?) + 1)
\end{lstlisting}
\end{subfigure}
\end{figure}

Regular expression capture groups {\color{blue}\si{\$2}} and
{\color{blue}\si{\$3}} match operands on the right-hand side to populate
\emph{read} facts, and the capture group {\color{blue}\si{\$1}} for the
left-hand side variable populates a \emph{write} fact. Some immediate problems
stand out in the regular expression approach. For example,

\begin{enumerate}
\item readability
suffers with escape sequences like \verb|\+| and repeated patterns like
{\color{blue}\verb|(\w)|} to match variables
\item patterns like
{\color{blue}\verb|\w+|} rely on overly simple assumptions on syntax that are
unlikely to generalize (e.g., to expressions)
\item regular expressions do not
natively expose metadata for source code (e.g., there's no easy way to
retrieve the location of matched variables, corresponding to
{\color{purple}\si{?}} in our facts). Such data would have to be accessed via an API
in a program script, and invites imperative implementation.
\end{enumerate}

Despite these shortcomings, the regular expression idea
prompts whether a more principled parsing approach (e.g., using PEGs~\citep{PEG})
could abstract the translation phase and better implement the idea. Our approach
pursues this line of thought. We use \ti{Comby}~\citep{vanTonderPPC}, a lightweight,
declarative, and language-aware parsing tool to implement the approach.
With \ti{Comby}, fact generation for our liveness example is expressed with a
declarative \emph{match template} (left) and \emph{rewrite template} (right):

\begin{figure}[ht]
\begin{subfigure}[t]{.30\columnwidth}
\begin{lstlisting}[basicstyle=\footnotesize\ttfamily,numbers=left,numbers=none,xleftmargin=6pt]
(?{\color{blue}\verb|$x|}?) = (?{\color{blue}\verb|$y|}?) + (?{\color{blue}\verb|$z|}?)
\end{lstlisting}
\end{subfigure}%
\begin{subfigure}[t]{.15\columnwidth}
\begin{lstlisting}[basicstyle=\footnotesize\ttfamily,numbers=left,numbers=none]
(?$\longrightarrow$?)
\end{lstlisting}
\end{subfigure}%
\begin{subfigure}[t]{.55\columnwidth}
\begin{lstlisting}[basicstyle=\footnotesize\ttfamily,numbers=none]
read((?{\color{blue}\verb|$y|}?), (?{\color{blue}\verb|$y.line|}?))
read((?{\color{blue}\verb|$z|}?), (?{\color{blue}\verb|$z.line|}?))
write((?{\color{blue}\verb|$x|}?), (?{\color{blue}\verb|$x.line|}?))
next((?{\color{blue}\verb|$x|}?), (?{\color{blue}\verb|$x.line|}?) + 1)
\end{lstlisting}
\end{subfigure}
\end{figure}

\ti{Comby} resolves some of the more glaring issues
%compared to
of the regular
expression counterpart:

\begin{enumerate}
\item no escape sequences and intuitive variable
metasyntax for binding matched values
\item parameterizable matching behavior for
lexical terms like variable identifiers or code block structures
\item built-in properties for matched values (e.g., {\color{blue}\si{\$x.line}} substitutes
{\color{blue}\si{\$x}}'s line number).
\end{enumerate}

An off-the-shelf Datalog engine like \ti{Souffl\'{e}}\footnote{\href{https://souffle-lang.github.io}{souffle-lang.github.io}} can consume generated these facts directly, allowing to easily express queries like
 \si{live(x,2)} to find which variables satisfying \si{x} are live at line 2, or
 \si{live("b",l)} to find all lines where variable \si{b} may be live.
A more complete liveness construction will involve more rules
for language-specific syntax; 
our example suffices to illustrate that domain-specific
rewriting for code structures and metadata present a workable solution for declarative fact
generation.
In the rest of this article we focus on call graph construction and
reachability, and show how the approach generalizes naturally to multiple
languages (Go, C, Zig) for this analysis domain.

We cover our approach for handling real-world programs
(\S~\ref{sec:approach}) and evaluate fact generation for call graphs on three
languages (\S~\ref{sec:eval}). We discuss related work in \S~\ref{sec:related}
and conclude in \S~\ref{sec:conclude}.

\section{Approach}
\label{sec:approach}

As a first
approximation, we pursue a purely syntax-driven approach. In this
approach, declarative templates specify syntactic patterns to match and
transform source code to Datalog facts. It is worth stating two caveats up
front. First, a purely syntax-driven approach cannot account for semantic
properties (e.g., type information) that may affect the domain of the analysis.
We treat this concern in more detail in \S~\ref{subsec:extend}. Second, our
declarative approach emphasizes ease of expressivity and trades precision when
interpreting a source input program. I.e., even at the syntactic level, we do
not strive to achieve parity with the exacting precision of language-specific
compilers or tooling for translating source code to facts. Since our approach is
a first step to proposing declarative fact generation, we focus on demonstrating
feasibility and surfacing challenges for practical use cases. We show that by
making these tradeoffs, a general method for declarative fact generation
\emph{can} make promising progress for declarative analysis, even across
multiple languages and syntaxes.

\subsection{Declarative syntax matching and rewriting}

We implement our approach with \ti{Comby},\footnote{\href{https://comby.dev}{\texttt{comby.dev}}} a tool to declaratively match and
rewrite source code syntax, and use it to translate relevant syntactic
constructs to Datalog facts. We choose \ti{Comby} because it provides flexible,
declarative abstractions for syntax rewriting while supporting multiple
languages. \ti{Comby} is language-aware and can correctly parse key program
features that allow easy execution of our core idea.

\ti{Comby} is especially suited for matching code blocks
 and disambiguating code from comments and strings. It provides
a mechanism to define custom match syntax and behavior.
The following custom \ti{Comby} definitions and syntax to implement
the approach in this article:

\begin{itemize}

\item {\color{blue}\si{\$x}} matches words like \si{hello} and
  contiguous well-balanced expression-like syntax like \si{(a + b)} or
  \si{print("hello world")}. It stops matching at whitespace boundaries and so does not match a string like
 \si{a + b}. It also does not match unbalanced code syntax like
  \si{foo)} in typical languages where expressions are expected to be well-balanced.

\item {\color{blue}\si{\$x*}} matches the same syntax as {\color{blue}\si{\$x}},
  but generalizes to match across whitespace and comments. It stops matching
  within the boundaries of a well-balanced code block or expression. For
  example, \verb|{|{\color{blue}\si{\$x*}}\verb|}| matches the body of the
  balanced braces and across new lines, irrespective of whitespace.

\item {\color{blue}\si{\$x?}} matches the same syntax as {\color{blue}\si{\$x}},
  but makes the match optional.

\item \si{"}{\color{blue}\si{\$x}}\si{"} matches the body of a well-quoted
  string. Unlike {\color{blue}\si{\$x}} (without quotes), the quoted variety
  implies that a data string may be any value, including balanced string values
  that contain unbalanced parentheses, like \si{"}item)\si{"}.

\end{itemize}

\subsection{Real-world application: Call graph reachability}

We demonstrate an end-to-end approach by targeting a call graph reachability
analysis. We consider three target languages for these analysis: Go, C, and Zig.
We use the \ti{Souffl\'{e}} Datalog framework~\citep{JordanSS16} to define and
run analyses, which consumes Datalog facts generated by \ti{Comby}. The rest of
this section explains capabilities and specification of declarative rewriting
with \ti{Comby}, followed by declarative Datalog definitions for computing call
graph reachability.

\begin{figure}[ht]
\begin{lstlisting}[basicstyle=\footnotesize\ttfamily,numbers=left,stepnumber=1,frame=single,xleftmargin=18pt,xrightmargin=8pt,frameround=ffft]
package main

import (?\verb|"|?)fmt(?\verb|"|?)

func one() int {
  return 1
}

func incr(n int) int {
  return n + one()
}

func main() {
  fmt.Printf((?\verb|"|?)%d: %d(?\verb|"|?), one(), incr(1))
}
\end{lstlisting}
\caption{\si{example.go}}
\label{fig:code}
\end{figure}

\begin{figure*}[t]
\begin{subfigure}[t]{\columnwidth}
\begin{lstlisting}[basicstyle=\footnotesize\ttfamily,numbers=none,style=comby,frame=single,xleftmargin=20pt,xrightmargin=20pt,frameround=tttt]
func @X$fX@(@Y...Y@) @X$r?X@ {@X$body*X@} @Y->Y@ @X$bodyX@

where nested, rewrite @X$bodyX@ {
  @X$cX@(@Y...Y@) @Y->Y@ edge((?\verb|"|?)@X$fX@(?\verb|"|?), (?\verb|"|?)@X$cX@(?\verb|"|?)).
}
\end{lstlisting}
\caption{Declarative source code rewriting to produce facts (EDB).}
\label{subfig:cg-edge}
\end{subfigure}
\begin{subfigure}[t]{\columnwidth}
\begin{lstlisting}[basicstyle=\footnotesize\ttfamily,numbers=left,numbers=none,frame=single,xleftmargin=15pt,xrightmargin=15pt]
.decl edge(x:symbol, y:symbol)
.decl calls(x:symbol, y:symbol)

calls(X,Y) :- edge(X,Y).
calls(X,Y) :- edge(X,K), calls(K,Y).
\end{lstlisting}
\caption{Call graph definitions for EDB facts and IDB relations in \ti{Souffl\'{e}}.}
\label{subfig:cg-decl}
\end{subfigure}
\caption{A wholly declarative specification to compute simple static call graph relations from Go code syntax.}
\label{fig:all-decl}
\end{figure*}

We demonstrate a declarative call graph construction with the Go program in Fig.~\ref{fig:code}.
Our objective is to emit facts that assert whether a function directly calls
another function. For example, we want to identify that function \si{incr}
contains a call site of function \si{one} on line 10. We represent this relation
with a fact \si{edge("incr", "one")}. The full specification to declaratively
rewrite source code to \si{edge} facts is shown in Fig.~\ref{subfig:cg-edge}
and comprises only four lines. We will walk through how this specification
operates shortly.

Once we obtain a set of all \si{edge} facts, we use Datalog definitions to
compute whether a function transitively calls (may reach) another function,
represented by the predicate \si{calls(X,Y)}. Fig.~\ref{subfig:cg-decl} defines
these rules in \ti{Souffl\'{e}}, which comprise only four lines. Armed with
initial facts, the definitions compute \si{calls} relations on-demand, or output
all \si{calls} facts, representing all relations in the entire call graph. In
practical terms, we can compute call graph reachability for any function (e.g.,
to discover dependencies) or process facts to generate a visual call graph.

We now explain the operation of the rewrite specification in
Fig.~\ref{subfig:cg-edge}. To generate \si{edge} facts, we start with the
following \ti{Comby} template to match all static functions:

\vspace{.5em}
\begin{lstlisting}[basicstyle=\footnotesize\ttfamily,numbers=none,style=comby,frame=single,frameround=tttt,xleftmargin=45pt,xrightmargin=45pt]
func @X$fX@(@Y...Y@) @X$r?X@ {@X$body*X@}
\end{lstlisting}
\vspace{.5em}

\noindent
The metavariable {\color{blue}\si{\$f}} matches the function identifier and
{\color{blue}\si{\$body*}} matches the function body within well-delimited braces
\verb|{|$\ldots$\verb|}|. The {\color{blue}\si{\$r?}} metavariable optionally
matches syntax that specify a return type (like \si{int} in our example).
Ellipses {\color{purple}$\ldots$} act as an anonymous variable matching the
function's parameters, which we don't use for call graph construction. All other
syntax matches concretely.

Next, we use a \ti{Comby} rule to match each call {\color{blue}\si{\$c}} \emph{within}
{\color{blue}\si{\$body}}, and emit the identifier for that call in a fact as
\si{edge(}\verb|"|{\color{blue}\si{\$f}}\verb|"|\si{,
}\verb|"|{\color{blue}\si{\$c}}\verb|"|\si{)}. The rule looks like this:

\vspace{.5em}
\begin{lstlisting}[basicstyle=\footnotesize\ttfamily,numbers=none,style=comby,frame=single,xleftmargin=35pt,xrightmargin=35pt,frameround=tttt]
where rewrite @X$bodyX@ {
 @X$cX@(@Y...Y@) @Y->Y@ edge((?\verb|"|?)@X$fX@(?\verb|"|?), (?\verb|"|?)@X$cX@(?\verb|"|?)).
}
\end{lstlisting}
\vspace{.5em}

\noindent
This \si{rewrite} rule overwrites {\color{blue}\si{\$body}} and appends a result
every time the pattern \si{{\color{blue}\$c}({\color{purple}...})} matches,
emitting an \si{edge} fact on a new line for each call site. By default
\ti{Comby} matches \si{{\color{blue}\$c}({\color{purple}...})} only to calls in
the top level \si{{\color{blue}\$body}}, and not nested calls. To handle nested
calls, we add an additional \si{nested} option to the rule that ensures the
rewrite rule fires for calls that nest, like the \si{one} call nested inside
\si{fmt.Printf} on line 14 of Fig~\ref{fig:code}. The value of
\si{{\color{blue}\$f}} is substituted for the function identifier in the match
template, and is in scope of the rewrite rule.

To output the full set of facts, we simply need to output
\si{{\color{blue}\$body}}, which we do by appending {\color{purple}\si{->
}}{\color{blue}\si{\$body}} to the match template. Putting this together, we
arrive at the complete specification in Fig.~\ref{subfig:cg-edge}. This emits
the desired facts for \si{example.go} in Fig~\ref{fig:code}.

\begin{lstlisting}[basicstyle=\footnotesize\ttfamily,numbers=none,xleftmargin=30pt,xrightmargin=30pt]
edge("incr", "one").
edge("main", "fmt.Printf").
edge("main", "one").
edge("main", "incr").
\end{lstlisting}

All together, we run one command line invocation of \ti{Comby} to generate these
facts, then a subsequent command line invocation of \ti{Souffl\'{e}} to consume them
and output the analyzed result. We can then query \ti{Souffl\'{e}} for the functions
reachable via \si{main} with the query \si{calls(main,X)} to yield the set of calls
\si{\{}\si{incr}, \si{one}, \si{fmt.Printf}\si{\}}.

\subsection{Complexities for declarative fact generation}
\label{subsec:extend}

Both syntactic and semantic features in modern languages impact the ability to
precisely recognize properties to encode in the Datalog domain. Taking call
graph reachability as an example, we discuss possibilities for extending
syntactic matching, and associated challenges thereof. We then expand on the
prospect of incorporating semantic information to overcome hurdles that inhibit
more precise declarative specification.

\nbf{Extending syntax matching.} We take Go as a representative modern language
that supports additional programming features that bear on call graph
construction, like methods and anonymous functions. At a syntactic level, we can
extend \ti{Comby} to match method definitions, e.g.,

\begin{lstlisting}[basicstyle=\footnotesize\ttfamily,numbers=none,style=comby,frame=single,frameround=tttt,xleftmargin=25pt,xrightmargin=25pt]
func (@X$vX@ @X$tX@) @X$fX@(@Y...Y@) @X$r?X@ {@X$body*X@}
\end{lstlisting}

\noindent
Which matches Go method syntax like:

\begin{lstlisting}[basicstyle=\footnotesize\ttfamily,numbers=none,style=comby,frame=single,frameround=ffff,xleftmargin=25pt,xrightmargin=15pt]
func (v Vertex) Abs() float64 {...}
\end{lstlisting}

We may continue to emit \si{edge} facts as-is for methods and ignore the
receiver type \si{\color{blue}\$t}. Alternatively, we could extend the domain to
distinguish calls via methods with a relation
\si{methodEdge("{\color{blue}\$t}", "{\color{blue}\$f}", "{\color{blue}\$c}")}. Whether to extend these definitions will
depend on the context of the practical application at hand. Notably, a
declarative approach can ease tailoring fact generation to particular applications, depending on context and language complexity.
On the other hand, we recognize that languages may use syntax that is awkward to match. An
attempt may even be fruitless for functional programs that make heavy use of
function passing or partial application. In these cases, we envision that
building on increasingly language-aware tooling\footnote{e.g., Tree-sitter,
\href{https://github.com/tree-sitter/tree-sitter}{\texttt{github.com/tree-sitter/tree-sitter}}.}
can enable more precise (but still declarative) methods for generating facts. We
observe that the benefit and tradeoff of our initial approach is
that simple and lightweight specifications (like those of
Fig.~\ref{subfig:cg-edge}) generalize well, if not precisely, to multiple
imperative languages like Go, Java, C, and so on. This attribute is compelling
where language-specific tooling is
absent or overly complex for the purpose at hand, as we show in~\S\ref{sec:eval}.

\nbf{Extending semantic matching.} Despite the tendency for languages like C to
call functions directly, syntax alone is generally not indicative enough to
precisely establish call relations. Consider the Go function in
Fig.~\ref{fig:calls.go}. Without knowing the context of imported packages and
locally scoped variables, it is syntactically ambiguous whether dot accesses
refer to local variables (like \si{p}) or imported packages (like \si{fmt}).
More generally, complex analyses like pointer-analysis benefit from type
information~\citep{aho2006compilers} and language-specific semantics will influence the specificity of
type constraints to generate facts.

\begin{figure}[h!]
\begin{lstlisting}[basicstyle=\footnotesize\ttfamily,numbers=left,stepnumber=1,frame=single,xleftmargin=30pt,xrightmargin=30pt,frameround=ffft]
import "fmt"

func main() {
	p := Printer{}
	p.Println("hello world")
	fmt.Println("hello world")
}
\end{lstlisting}
\vspace{-1em}
\caption{}
\label{fig:calls.go}
\end{figure}

Due to language-specific semantics, we must appeal to tooling in e.g., compilers
to resolve ambiguity. Recent advances use language servers~\citep{graal-lsp} that
expose semantic properties via an API. Our idea is to incorporate this semantic
information via language servers, where declarative specification is
\emph{decoupled} from advanced, external processing that provides contextual semantic
information. A prototype query that integrates \ti{Comby} with the Go language
server, for example, could allow conditionally emitting \si{edge} facts where
the identifier refers to an imported package:\footnote{We use a convention of
\texttt{package} to refer to a type in the type environment---a fully general
solution relies on a server maintaining this state, and appropriate
conventions to refer unambiguously to types.}

\begin{figure}[h!]
\begin{lstlisting}[basicstyle=\footnotesize\ttfamily,numbers=none,style=comby,frame=single,xleftmargin=30pt,xrightmargin=30pt,frameround=tttt]
where rewrite @X$bodyX@ {
 @X$cX@(@Y...Y@) @Y->Y@
   @X$cX@.type == (?\verb|"|?)package(?\verb|"|?),
   edge((?\verb|"|?)@X$fX@(?\verb|"|?), (?\verb|"|?)@X$cX@(?\verb|"|?)).
}
\end{lstlisting}
\end{figure}

\begin{figure*}[t!]
\includegraphics[scale=0.156]{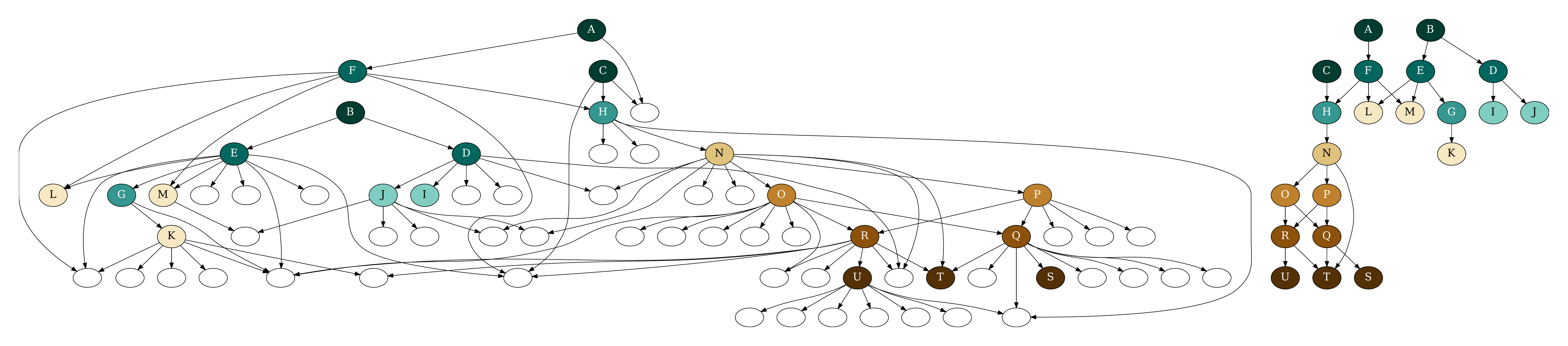}
\caption{Two call graphs generated for the \ti{upgrade} package in the
  \ti{syncthing} Go project. Nodes correspond to functions, with shortened
  labels for visual comparison. On the left, the visual call graph for edges generated
  by the declarative specification in Fig.~\ref{subfig:cg-edge} (our approach).
  On the right, the static call graph generated by existing tool \ti{go-callvis}
  starting at root functions in the same package. The takeaway is that our facts
  yield a subgraph that is isomorphic to the static call graph generated by the
  language-specific tool \ti{go-callvis} (i.e., no edges are missed by our
  approach). Note: as a matter of design, \ti{go-callvis} omits visualizing
  certain call sites by default (e.g., built-in functions like \ti{len}). We
  do report these in our approach, which accounts for unlabeled nodes in the left
  graph.}
\label{fig:compare-cg}
\end{figure*}

\section{Evaluation}
\label{sec:eval}

Our evaluation considers the feasibility, accuracy, and speed of declaratively
emitting call facts on large real-world projects written in three languages: Go,
C, and Zig.
We chose Go because it is a popular language with existing tooling to
generate and visualize call graphs. This tooling provides a basis for a
qualitative comparison in~\S\ref{sec:eval-qual}. We chose Zig because it is a
relatively new language that further demonstrates the generality and speed of
our declarative approach. To demonstrate the approach at scale, we run fact
generation over the Linux kernel, which is written in C. We show that we can quickly and
easily generate call graph relations without language-specific in
tooling~\S\ref{sec:eval-speed}. All experiments were performed on a 6-core
desktop machine (Core i5-9400F CPU, Ubuntu 20.04 LTS, 16GB RAM).
All data and tooling is released toward Open Science and available at
\href{https://github.com/comby-tools/direct-to-datalog}{\si{github.com/comby-tools/direct-to-datalog}}
and archived by citable DOI \href{https://zenodo.org/record/5520885#.YUvYR7hKjuo}{\si{10.5281/zenodo.5520885}}.

\subsection{Qualitative call graph comparison}
\label{sec:eval-qual}

We qualitatively compare our call graph output to that of \texttt{go-callvis},\footnote{\href{https://github.com/ofabry/go-callvis}{\ti{https://github.com/ofabry/go-callvis}}} a Go-specific tool. Our objective is to compare how well our syntactic approach
recovers static call relations, and we use \texttt{go-callvis} output as ground
truth. \texttt{go-callvis} fully parses Go code and qualifies functions and
methods by package imports. Its final call graph representation is therefore
richer than what our syntax-driven approach supports. Thus, we use
\texttt{go-callvis} to qualify how well our approach approximates ground truth,
not whether we can achieve tooling parity. Fundamentally, we are comparing how
well the four-line specification in Fig.~\ref{subfig:cg-edge} approximates a
relevant subset of call edges that result from language-specific tooling and
libraries that span thousands of lines of code.

For a tractable qualitative comparison, we generate call graphs for the
\ti{upgrade} package in the \ti{syncthing} Go project.\footnote{The \ti{upgrade}
package is the canonical example visualized in the
\href{https://github.com/ofabry/go-callvis\#examples}{\ti{go-callvis}
  repository}.} The package comprises approximately 600 lines of Go code across
5 files. Fig.~\ref{fig:compare-cg} visualizes the static calls found in our
approach (left) versus \ti{go-callvis} output (right). To ease visual
comparison, we show \ti{go-callvis} output rooted at functions declared in the
\ti{upgrade} package, and do not render functions that call into the
\ti{upgrade} package from external packages. Note that our approach reports more
nodes and edges because \ti{go-callvis}, by design, omits certain functions in
the Go standard library by default.\footnote{It is also possible to similarly
tailor rules in \ti{Comby} to omit output of such edges; we elide discussion for
brevity.}

The key result is that our approach yields a subgraph that is
isomorphic to the \ti{go-callvis} static call graph. That is, \emph{no edges are
missed} by our approach.
This result
supports the premise that succinct, declarative patterns can accurately generate facts to
drive
analyses, to, e.g., compute reachability properties over static call
graphs.

\subsection{Fact generation speed and generality}
\label{sec:eval-speed}

Table~\ref{tab:results} demonstrates the speed of our approach on
real-world repositories. To demonstrate generality of our approach we evaluate
over large projects for Go and C, as well as projects for a trending language
called Zig. Minimal changes are needed to adapt the Go pattern in
Fig.~\ref{subfig:cg-edge} to work for static C and Zig calls. Due to Zig's
C-like syntax, we only need to change the function keyword to match \ti{fn}:

\vspace{.5em}
\begin{lstlisting}[basicstyle=\footnotesize\ttfamily,numbers=none,style=comby,frame=single,xleftmargin=20pt,xrightmargin=20pt,frameround=tttt]
fn @X$fX@(@Y...Y@) @X$r?X@ {@X$body*X@} @Y->Y@ @X$bodyX@
\end{lstlisting}
\vspace{.5em}

\noindent Similarly, we support C syntax with:

\vspace{.5em}
\begin{lstlisting}[basicstyle=\footnotesize\ttfamily,numbers=none,style=comby,frame=single,xleftmargin=20pt,xrightmargin=20pt,frameround=tttt]
@X$fX@(@Y...Y@) {@X$body*X@} @Y->Y@ @X$bodyX@
\end{lstlisting}
\vspace{.5em}

Table~\ref{tab:results} shows that our approach is fast on very large projects.
Fact generation for some of the largest Go projects finishes in under 3 minutes.
At the extreme, we generate over 3.5 million facts over roughly 21 million lines
of C code in 40 minutes for the Linux kernel. We successfully generate thousands
of facts over all three languages. Interestingly, we observe
a consistency in the ratio of call facts to functions: on average, across all
projects, we generate 9.39 call facts per function (standard deviation $\approx$1.27). This is a promising positive indicator that our syntax patterns generate
facts consistently across languages.\footnote{Dually, this observation raises the
prospect of cross-language fact generation for
``natural'' software properties~\citep{naturalness}.}

\begin{table}
\footnotesize
\centering
\begin{tabular}{llrrrr}
\toprule
\bf Lang         & \bf Project   & \bf KLOC        & \bf \# Facts         & \bf \# Funcs  & \bf Time \\
\midrule
\mr{3}{Go}       & Go            & 1,701           & 321,084              & 34,658            & 2m56s    \\
                 & K8s           & 2,436           & 334,308              & 31,669            & 2m33s    \\
                 & Sync          & 131             & 21,665               & 1,908             &    9s    \\
\midrule
\mr{3}{Zig}      & Zig           & 467             & 70,461               & 7,414             &   42s    \\
                 & ZLS           & 16              & 2,454                & 256               &    2s    \\
                 & Dida          & 5               & 1,381                & 165               &    2s    \\
\midrule
\mr{1}{C}        & Linux         & 20,916          & 3,660,511            & 513,264           & 39m14s  \\
\bottomrule
\end{tabular}
\caption{Call graph fact generation over Go, Zig, and C projects. \underline{KLOC} is the thousands of Lines of Code processed (1,000 KLOC is 1 million lines of code). \underline{\# Facts} is the number of static calls edges generated. \underline{\# Funcs} is the number of functions matched.
%\underline{Time} is the wall clock time.
In general, fact generation is fast and yields many static call relations across all languages.}
\label{tab:results}
\end{table}

\section{Related Work}
\label{sec:related}

Pioneering work in declarative program analysis demonstrates efficient and
succinct formulations using
Datalog~\citep{WhaleyACL05,JordanSS16,BravenboerS09}. Existing work focuses on
deep, language-specific properties and domains (e.g., pointer analysis for
Java). These correspondingly implement language-specific frameworks and parsing
routines to extract facts in the domain. None, to our knowledge, have attempted
to generalize a declarative approach for fact generation across multiple
languages, nor evaluate the feasibility of fact generation for cross-language
properties like call graph construction. Alternative tools and frameworks exist
for declaratively transforming syntax in multiple
languages~\citep{srcml,Cordy06,workbenches}. We are not aware of any that
attempt to fill the gap for declaratively generating Datalog facts. We used
\ti{Comby}~\citep{vanTonderPPC} because it is simple, language-accessible, and
fast; existing tools with similar properties may also implement
the fact generation routines described in this article.

% GNU cflow

\section{Conclusion}
\label{sec:conclude}

We presented the first approach that investigates the feasibility and appeal of
an end-to-end declarative pipeline across multiple languages, where declarative
code rewriting directly outputs Datalog facts. Our key result shows how we use
this approach to generate thousands of static call edge facts across
multiple languages (Go, C, and Zig) and that it scales to large, real world
projects. The declarative specification requires less than 10 lines of code and
can achieve a degree of qualitative parity with language-specific call graph
tools implemented in hundreds of lines of imperative code
(\S~\ref{sec:eval-qual}). Datalog engines like \ti{Souffl\'{e}} directly consume facts
generated by our approach and can then answer, e.g., call graph reachability
properties. We envision that our syntax-driven approach can incorporate
language-specific semantic information (via \cite{graal-lsp}) to expand
analysis kinds and precision (e.g., pointer analysis) while retaining the
benefits of declarative specification.

\section*{Acknowledgments}
\noindent
This study was conducted independently by the author and received no
funding support. No funding support is implied by the author's affiliation to
Sourcegraph, Inc. by this manuscript.

%% The Appendices part is started with the command \appendix;
%% appendix sections are then done as normal sections
%% \appendix

%% \section{}
%% \label{}

%% References
%%
%% Following citation commands can be used in the body text:
%% Usage of \cite is as follows:
%%   \cite{key}         ==>>  [#]
%%   \cite[chap. 2]{key} ==>> [#, chap. 2]
%%

%% References with BibTeX database:
% \balance

% There are no strict requirements on reference formatting at submission. References can be in any style or format as long as the style is consistent. Where applicable, author(s) name(s), journal title/book title, chapter title/article title, year of publication, volume number/book chapter and the article number or pagination must be present. Use of DOI is highly encouraged. The reference style used by the journal will be applied to the accepted article by Elsevier at the proof stage.
% https://www.elsevier.com/journals/journal-of-systems-and-software/0164-1212/guide-for-authors

\balance
\bibliographystyle{elsarticle-harv}
\bibliography{misc}

%% Authors are advised to use a BibTeX database file for their reference list.
%% The provided style file elsarticle-num.bst formats references in the required Procedia style

%% For references without a BibTeX database:

% \begin{thebibliography}{00}

%% \bibitem must have the following form:
%%   \bibitem{key}...
%%

% \bibitem{}

% \end{thebibliography}

\end{document}